\documentclass{JHEP3}
\usepackage{graphics}
\usepackage{amssymb}
\usepackage{epsfig}

\title{Discrete Gauge Symmetries, Baryon Number and Large Extra Dimensions} 
\author{Andrew Pawl\\ Michigan Center for Theoretical Physics,
University of Michigan, Ann Arbor, MI, USA\\ 
E-mail: \email{apawl@umich.edu}}

\abstract{Krauss and Wilczek have
shown that an unbroken discrete gauge symmetry 
is respected by gravitationally mediated processes.  This has led to
a search for such a symmetry compatible with the standard model or
MSSM that would protect protons from gravitationally mediated decay
in a universe with a low scale for quantum gravity
(large extra dimensions). The fact that the discrete symmetry
must remain unbroken and have a gauge origin puts important
restrictions on the space of possible discrete symmetries.} 

\preprint{MCTP-04-19}
\keywords{eld, dfs, blh}

\begin{document}

\section{Introduction}

Gravitationally mediated proton decay has long been postulated to result
from the ``spacetime foam'' predicted in the current understanding of
quantum gravity \cite{Wheeler,Hawking0,Hawking}.  
This process is completely negligible
in traditional cosmologies, with the Planck scale safely relegated to
extremely high energies ($M_{pl} \sim 10^{19}$ GeV).  It has been 
proposed, however, that in a cosmology with 
the quantum gravity scale reduced by the presence of
large extra dimensions \cite{hamed},
 proton decay can proceed extremely rapidly \cite{gordy}.

One obvious way to circumvent this issue is to invent a new gauge
symmetry that protects against effective operators that mediate fast
proton decay.  An unbroken gauge symmetry, however, would require a new massless
gauge boson, which is phenomenologically unacceptable.  A
broken symmetry, on the other hand, is known to be violated by
black hole physics in the classical regime
\cite{beck}.  For this reason, it is often assumed that the virtual 
Planck-scale 
black holes arising in a spacetime foam picture will similarly violate
broken gauge symmetries \cite{Hawking2,worm}.  Thus, we are left
with no way to protect protons against decay mediated by virtual black
holes.

This difficulty was resolved by Krauss and Wilczek 
\cite{krauss} who proposed that a surviving discrete remnant of a
broken U(1) gauge symmetry could protect protons while at the same time
allowing the associated gauge boson to gain a large mass.  
They showed that a gauge symmetry broken to a discrete
remnant will still have an
associated conserved surface integral which could be used to detect
the presence of charge swallowed by a black hole.

Our purpose in this paper is to decide if there are any phenomenologically
acceptable discrete gauge symmetries (arising from a U(1) gauge symmetry)
that can protect against fast proton
decay and remain unbroken in the low energy standard model or
MSSM.  We wish to proceed in a manner analogous to  
Ib\'a\~nez 
and Ross's work \cite{ross}, in which they constructed a catalog of discrete symmetries that
protect against SUSY-mediated proton decay.

\section{The Setting}

Classical black holes fail to respect a broken gauge symmetry regardless
of the energy scale that characterizes the breaking \cite{beck}.  By
analogy, therefore, it has been assumed that quantum black holes also
fail to respect any broken symmetry \cite{Hawking,Hawking2,worm}.  
By exploiting this argument, Hawking, Page and Pope showed
that a spacetime foam of
virtual black holes can lead to an effective four-point interaction
converting two quarks into a lepton plus antiquark \cite{Hawking}.
This interaction will be suppressed by inverse powers of the Planck
mass, so that for $M_{pl} = 10^{19}$ GeV, the predicted rate of
proton decay is well below experimental bounds.  These authors
also examined the effect of spacetime foam on the masses of particles. 
Their results
indicate that quantum gravity effects will not
produce fermion and vector boson 
masses.  Scalars, on the other hand, were shown to acquire a mass
of order the Planck mass, which poses
a problem for the Higgs mechanism of electroweak symmetry breaking.

More recently, with the advent of theories with a low scale for quantum
gravity \cite{hamed}, the spacetime foam picture been revisited.
In models with one or more large extra dimensions, it is a desired feature
that the quantum gravity scale be nearly equal to the scale
for electroweak symmetry breaking so as to solve the hierarchy
problem.  This device eliminates the problem of large scalar masses
in the spacetime foam picture as well, since the contributions
from gravitational effects are reduced along with
the Planck scale.  This scenario is not completely
beneficial, however, since lowering the Planck scale only
serves to make the
four-point operators which produce proton decay more
important (less suppressed).
Calculations show that the proton lifetime can be a very important
constraint on theories with large extra dimensions if we assume 
the spacetime foam picture is correct.  In fact,
the current experimental bounds on proton decay would require
the scale for quantum gravity to be high enough so as to be
unreachable in any current or proposed collider \cite{gordy}.

For this reason, it is of interest to find a discrete
symmetry that forbids proton decay (or at least the
dimension six operators that result in fast proton decay), 
can remain
unbroken at low energies, and could potentially arise as
the remnant of a gauged symmetry.  All three conditions are
necessary to protect protons
in the spacetime foam picture with a low scale
for quantum gravity.
This leads to the constraint that the electroweak Higgs(es)
must
be uncharged with respect to any discrete gauge symmetry that is to
protect protons against decay.  Otherwise, the symmetry is broken today
and its protection is lost. 

If we want to retain the Higgs mechanism for fermion masses, an uncharged
Higgs leads directly to a constraint on the discrete charges of the
known fermions.
We would like to implement these constraints on the Ib\'a\~nez-Ross (IR)
catalog of discrete symmetries \cite{ross}.  Unfortunately, their work
assumes that hypercharge is unbroken.  This is unacceptable if we are
to find a symmetry that remains unbroken after the electroweak Higgs
gets a VEV.  We will, however, borrow from their terminology.  
In the IR language, we can easily construct the constraints that
apply to a discrete symmetry that is unbroken today.  The
first constraint is provided by fermion mass terms:
\begin{eqnarray}
\label{eq:uquark}
	\alpha_{Q_{i}} + \alpha_{u_{i}} = 0 \: {\rm mod} \: N\\
\label{eq:dquark}
	\alpha_{Q_{i}} + \alpha_{d_{i}} = 0\:{\rm mod} \: N  \\
\label{eq:lepton}
	\alpha_{L_{i}} + \alpha_{e_{i}} = 0\: {\rm  mod} \: N \\
\label{eq:neut}
	\alpha_{L_{i}} + \alpha_{\nu_{i}} = 0 \: {\rm mod}\: N
\end{eqnarray}
where each $\alpha$ represents the integer charge of the (super)field
noted under the discrete gauge group of interest and $i$ is
a family index (ranging from one to three). 

Before moving on, we should comment on neutrino masses.
 Note that relation
(\ref{eq:neut}) would be necessitated by Dirac masses.  Many
of the popular
scenarios for neutrino mass in a universe with
 large extra dimensions utilize Dirac neutrinos
\cite{original}.  It is, however, possible that 
neutrinos possess only Majorana mass terms.  In effect, this
case is actually more restrictive.  Equation (\ref{eq:lepton}) is
unmodified.  Further, we would be forced to constrain $\alpha_{L}$ to equal
$N/2$ to allow Majorana terms without further breaking our
$\mathbb{Z}_{N}$ symmetry.  As we shall see, the conditions 
above are already sufficient to place significant constraints
on the space of allowed $\mathbb{Z}_{N}$ symmetries, and so
we will not assume neutrino Majorana masses are allowed.  

We can express the restrictions of equations (\ref{eq:uquark})-(\ref{eq:neut}) 
very succinctly (again, following the terminology of \cite{ross}):
\begin{equation}
      \vec{\alpha}_{i} = (\alpha_{Q_i},-\alpha_{Q_i},-\alpha_{Q_i},\alpha_{L_i},-\alpha_{L_i},
       -\alpha_{L_i})
\end{equation}
where $\vec{\alpha}_{i}$ is a shorthand way of denoting the charges of
each family of (super)fields under the action of a discrete symmetry.
By definition:
\begin{equation}
	\vec{\alpha}_{i} \equiv (\alpha_{Q_{i}}, \alpha_{u_{i}}, \alpha_{d_{i}},
	\alpha_{L_{i}}, \alpha_{e_{i}}, \alpha_{\nu_{i}})
\end{equation}
(note that we have broken from \cite{ross} by eliminating 
$\alpha_{H}$ (since it has been set to zero), and by
adding $\alpha_{\nu}$ and a family index $i$).

This is a dramatic restriction, as it means the discrete symmetries 
which can remain unbroken today are parameterized by six integers (assuming
three families):
$\alpha_{Q_{i}}$ and $\alpha_{L_{i}}$.  
In fact, we can be even more restrictive by using the experimental
observations that quarks and neutrinos mix among the families. 
This implies that we can do away with the
family index.  Any symmetry that remains unbroken today must 
be family independent.  (Family-dependent symmetries are often
used in model building to give approximately correct CKM mixing or
neutrino mixing parameters.  Such models, however,
rely on the presence of one or more Higgs-type fields that are
charged under the symmetry.  We can have no such Higgs without breaking
our symmetry and losing its protection.)  Thus, we have limited the possible
discrete symmetries to those under which all families have the
charges:
\begin{equation}
	 \vec{\alpha} = (\alpha_{Q},-\alpha_{Q},-\alpha_{Q},\alpha_{L},-\alpha_{L},-\alpha_{L}).
\end{equation}

\section{Forbidding Fast Baryon Number Violation}
\label{sec:cons}

To forbid fast proton decay, our discrete symmetry must forbid the
operators $uude$, $QQQL$ and (perhaps) $udd\nu$.  Each of these results in
the same constraint on the charges $\alpha_{Q}$ and $\alpha_{L}$:
\begin{equation}
\label{eq:protdec}
	3\alpha_{Q} + \alpha_{L} \ne 0  \: {\rm mod}\: N.
\end{equation}

The above condition is all that is required to forbid fast 
gravitationally mediated
proton decay.  
There is, however, one additional signature of
baryon violation which has been experimentally investigated.  That
process is neutron-antineutron oscillation, which changes the
baryon number of the universe by $\Delta B = 2$.
Such oscillations would be mediated by effective six-quark operators of the
form $uddudd$.  

The experimentally observed lack of $\Delta B =2$
processes can also serve as an important constraint on models
with low Planck scale \cite{gordy}.
Current limits would require
that the scale for quantum gravity
be restricted to $M_{qg} > 10^{5} {\rm GeV}$ \cite{sacha}.  Thus,
if we hope to observe the effects of quantum gravity at the LHC
or indeed to solve the hierarchy problem with minimal
fine-tuning, we must forbid neutron-antineutron oscillation
as well as proton decay. 
To accomplish this, we must impose the 
inequality:
\begin{equation}
\label{eq:neu-antineu}
	6\alpha_{Q} \ne 0  \: {\rm mod}\: N.
\end{equation}

Equations (\ref{eq:protdec}) and (\ref{eq:neu-antineu}) are
important constraints.  Equation (\ref{eq:neu-antineu}) alone
eliminates any 
unbroken $\mathbb{Z}_{N}$ symmetry with $N=2, 3$ or 6 
from consideration as a means to prevent fast baryon
number violation in a universe with a low scale for quantum
gravity.  It
also eliminates the possibility that $\alpha_{Q} = 0$.  
Adding the constraint of equation (\ref{eq:protdec})
leaves only a handful of possibilities with $N \le 6$. 
These are listed
in Table \ref{tab:0}.

\TABLE{\begin{tabular}{|l|l|l|}
        \hline $N$  &  $\alpha_{Q}$ & $\alpha_{L}$ \\\hline
       4   &  1 & 0   \\
       4   &  1 & 2   \\
       4   &  1 & -1  \\
       5   &  1 &  0  \\
       5   &  1 &  1  \\
       5   &  1 &  -2  \\
       5   &  1 &  -1  \\
       5   &  2 & 0   \\
       5   &  2 & 1  \\
       5   &  2 & 2  \\
       5   &  2 & -2  \\\hline
\end{tabular}
\caption{Charges $\alpha_{Q}$ and $\alpha_{L}$ under
all possible (independent) unbroken $\mathbb{Z}_{N}$ symmetries with
$N \le 6$ which
satisfy equations (\ref{eq:protdec}) and (\ref{eq:neu-antineu}).}
\label{tab:0}}

\section{Anomalies}

\label{sec:anomalies}

Further constraints on the charges $\alpha_{Q}$ and $\alpha_{L}$ 
could be obtained
by considering discrete gauge anomalies \cite{anom,ibanez}.
Unless the anomalies cancel, it is impossible for the discrete 
symmetry to be a remnant
of a gauged U(1).  

Unfortunately, discrete anomalies are notoriously ambiguous
\cite{dine,ibanez}.  Theoretically, we should have several
anomalies to consider.  The U(1) giving rise to our discrete
symmetry (henceforth ${\rm U_{D}(1)}$) must be compatible with the
SU(3), SU(2) and ${\rm U_{Y}(1)}$ of the standard model, giving
rise to possible (nontrivial) anomalies of the form: $\mathbb{Z}_{N}^{3} $,
$\mathbb{Z}_{N}^{2} \times {\rm U_{Y}(1)}$, 
$\mathbb{Z}_{N} \times {\rm U_{Y}(1) \times U_{Y}(1)}$,
$\mathbb{Z}_{N} \times {\rm SU(2) \times SU(2)}$,
$\mathbb{Z}_{N} \times {\rm SU(3) \times SU(3)}$, and the
mixed gravitational anomaly.  We will consider each
in turn.

The $\mathbb{Z}_{N}^{3}$ is basically non-constraining due to the possible
existence of heavy fermions that are fractionally charged under the
remnant $\mathbb{Z}_{N}$ discrete symmetry \cite{dine}.

The mixed hypercharge anomalies are similarly difficult to
evaluate due to 
the unknown relative normalizations
of the ${\rm U_{D}(1)}$ and ${\rm U_{Y}(1)}$ groups
\cite{anom}.  

The mixed SU(3) anomaly is trivial.  Without 
exotic fermions it 
is proportional to:
\begin{equation}
\label{eq:su3}
    2\times 3 \alpha_{Q} + 3\alpha_{u} + 3\alpha_{d} = 6\alpha_{Q}-6\alpha_{Q} = 0  
\end{equation}	
so that the anomaly cancels for any choice of $\alpha_{Q}$ and $\alpha_{L}$.

The gravitational anomaly is similarly trivial.

The most promising anomaly constraint is provided by the
mixed SU(2) anomaly.  Here, extremely massive fermions
cannot contribute because the symmetry protects the masses.
We will therefore assume that the only fermions charged under
the weak SU(2) symmetry of the standard model are
are the three known families of quarks and leptons.
In that case, the discrete 
$\mathbb{Z}_{N}\times {\rm SU(2)} \times {\rm SU(2)}$
will be proportional (mod $N$) to:
\begin{equation}
     3\times3 \alpha_{Q} + 3 \alpha_{L} = 9\alpha_{Q}+3\alpha_{L}.
\end{equation} 
Direct cancellation of the SU(2) anomaly would then require:
\begin{equation}
\label{eq:anomaly}
9\alpha_{Q}+3\alpha_{L} = 0 \: {\rm mod} \: N.
\end{equation} 
There is, however, one other 
possibility to consider.

\TABLE{\begin{tabular}{|l|l|}
        \hline $p$  &  Allowed N \\\hline
      1   &  3  \\
        2         &  3, 6  \\
       3         &  9  \\
       4         &  6, 12  \\
       5         &  15  \\
       6         &  9, 18  \\
       7         &  21  \\
       8         &  12, 24  \\
       9         &  27  \\
      10         &  15, 30 \\\hline
\end{tabular}
\caption{Values of $N$ which satisfy
the constraint of equation (\ref{eq:anomaly}) for
the given values of $p \equiv 3\alpha_{Q} + \alpha_{L} \:{\rm mod} \: N$.}
\label{tab:1}}

Anomaly cancellation can also be achieved via a version of the
Green-Schwarz (GS) mechanism
\cite{gs}.  In heterotic string theory, the GS mechanism is 
basically model independent, since it involves only the dilaton superfield.
In type II string theories
with D-branes and orientifolds (or type I string theories) which
can accommodate a low string scale, on the other hand, the GS
mechanism is generalized because of the presence
of Ramond-Ramond (RR) 2-forms \cite{IRU98}.  Thus, if we construct
such a theory which involves $M$ RR 2-forms $B^{i}$ ($i = 1,2,...,M$)
we may expect terms in the Lagrangian of the form \cite{AFIRU00}:
\begin{equation}
\label{eq:gslagrange}
	\mathcal{L}_{GS} = \sum_{i=1}^{M} c_{i} B^{i}\wedge F_{D} + d_{i} \eta^{i}
        \; {\rm tr}\left(F_{\rm SU(2)} \wedge F_{\rm SU(2)}\right)
\end{equation}
where $\eta^{i}$ is the scalar dual to $B^{i}$, $F_{D}$ is the field strength
associated with the U$_{D}$(1) gauge boson, and $F_{\rm SU(2)}$ is the
field strength associated with the SU(2) group (we mention only the SU(2) group
because it alone yields a nontrivial mixed U$_{D}$(1) anomaly -- the
couplings can be present for any group).  The model-dependent coefficients 
$c_{i}$ and $d_{i}$ will then make a contribution to the mixed 
U$_{D}$(1)$\times$SU(2)$\times$SU(2) anomaly \cite{IMR01}:
\begin{equation}
	\delta A = \sum_{i=1}^{M} c_{i} d_{i}.
\end{equation}
Currently, there are no generic constraints on the values of the
coefficients $c_{i}$ and $d_{i}$.  Thus, if the GS mechanism
is applicable, it removes 
the constraint of equation
(\ref{eq:anomaly}).  

It is important to remember, however, that the GS mechanism apparently breaks
the associated anomalous U(1) symmetry down to
a global symmetry \cite{gs,BIQ98}. 
This has been used as a means to
protect protons in string theory \cite{IQ99}, but in the case of
virtual black hole mediated decay a residual global
U(1) is not
sufficient protection.  Instead, the gauge symmetry must be
broken in a way that still admits nontrivial strings
or vortex solutions \cite{krauss,coleman}.  The GS mechanism
breaks the associated U(1) via the St\"uckelberg mechanism \cite{GIIQ02},
which does not obviously admit such strings.  Of course, this
is hardly conclusive.  It has recently been
asserted that the GS mechanism does not prohibit string configurations, 
and postulated that all
discrete symmetries in string constructions are gauged discrete symmetries
\cite{newdine}.  For the purposes of this work, we will consider this
an open question.  We therefore catalog the constraints that would
become available if it is shown that the GS mechanism \emph{does not} give
rise to gauged discrete remnants.  If the GS mechanism is shown to allow
discrete gauge symmetries, we are left with the (still quite restrictive)
constraints given in Section \ref{sec:cons}.

\TABLE{\begin{tabular}{|l|l|l|l|}
\hline     $N$ &     $\alpha_{Q}$    &    $\alpha_{L}$  &  $p$  \\\hline
     9 &      1    &    0  &  3  \\
     9 &      1    &    3  &  6  \\
     9 &      2    &    0  &  6  \\
     9 &      2    &    6  &  3  \\
     9 &      4    &    0  &  3  \\
     9 &      4    &    3  &  6  \\\hline
\end{tabular}
\caption{Independent combinations of $\alpha_{Q}$ and $\alpha_{L}$ allowed
by constraints (\ref{eq:protdec}),
(\ref{eq:neu-antineu}) and (\ref{eq:anomaly}) for $N=9$ (the only allowed $N \le 10$).}
\label{tab:2}}

In the case that the GS mechanism is not available, we must 
impose the constraints of equations (\ref{eq:protdec}),
(\ref{eq:neu-antineu}) and (\ref{eq:anomaly}) to decide what values
of $\alpha_{Q}$ and $\alpha_{L}$ are allowed for a given $\mathbb{Z}_{N}$ symmetry.
For convenience, we now define the quantity
$p \equiv 3\alpha_{Q}+\alpha_{L}$.  With this definition,
equation (\ref{eq:protdec}) requires 
$p \ne 0 \: {\rm mod} \: N$
and equation (\ref{eq:anomaly}) becomes $3p = 0 \: {\rm mod} \: N$.
These two equations for $p$ allow us to find the allowed $N$ values 
corresponding to any given $p$.  Table \ref{tab:1} shows the
possible $N$ values for $p$'s up to 10.  
We next add the constraint of equation (\ref{eq:neu-antineu}).
In table \ref{tab:2}, we report the only (independent)
$\alpha_{Q}$ and $\alpha_{L}$ charges which satisfy all three
constraint equations for $N \le 10$.
This table
reports the only possible (independent) $\mathbb{Z}_{N}$ discrete symmetries
with $N \le 10$ which will remain unbroken and forbid gravitationally
mediated proton decay and gravitationally mediated neutron-antineutron
oscillation in the case that the GS mechanism is \emph{unavailable} for anomaly
cancellation.  Note that under our assumptions -- three families of quarks/leptons and
Dirac mass terms for neutrinos (recall that Majorana masses tend to be \emph{more}
restrictive) --
 any $\mathbb{Z}_{N}$ arising from a U(1) gauge symmetry
 which is to remain unbroken
and protect against fast gravitationally-mediated
baryon violation in a universe with a low
scale for quantum gravity must have $N \ge 9$ if the GS mechanism is
unavailable.  

\section{Conclusions}

If we assume by analogy with classical physics
that broken symmetries are incapable of protecting
against gravitationally mediated proton decay
and neutron-antineutron oscillation, we are led to
the constraint that the electroweak Higgs(es) must be neutral
under any surviving discrete gauge symmetry that would be used
to protect baryon number.  This simple constraint, when 
coupled with the assumption of three families of quarks
and leptons and Dirac neutrino masses,
is sufficient
to eliminate all $\mathbb{Z}_{N}$ symmetries
arising from a gauged U(1) with $N = 2,3$ or 6 from consideration.
Perhaps more importantly, unbroken $\mathbb{Z}_{N}$ symmetries
are parameterized by only two independent charges. 

\section*{Acknowledgments}

I wish to thank G. Kane for suggesting the topic.  I am indebted to
referees for pointing out mistakes in earlier drafts and for
helpful and important comments.
Discussions with F. Adams and
M. Perry are gratefully acknowledged.


\begin{thebibliography}{00}

\bibitem{Wheeler} J.A. Wheeler, in {\it Relativity, Groups and Topology},
ed. B.S. DeWitt and C. DeWitt-Morette, Gordon and Breach, 1964.

\bibitem{Hawking0} S.W. Hawking, \npb{144}{1978}{349}.

\bibitem{Hawking} S.W. Hawking, D.N. Page and C.N. Pope,
\plb{86}{1979}{175}.

\bibitem{hamed} N. Arkani-Hamed, S. Dimopoulos and G.R. Dvali,
\plb{429}{1998}{263}; \prd{59}{1999}{086004};
I. Antoniadis, N. Arkani-Hamed, S. Dimopoulos and G.R. Dvali,
\plb{436}{1998}{257}.

\bibitem{gordy} F.C. Adams, G.L. Kane, M. Mbonye and M.J. Perry,
\ijmpa{16}{2001}{2399}.

\bibitem{beck} J.D. Bekenstein, \prd{5}{1972}{1239}, 2403;
C. Teitelboim, {\it ibid.}
2941; R.H. Price, {\it ibid.} 2419,
2439.

\bibitem{Hawking2} S.W. Hawking, \prd{14}{1976}{2460}.

\bibitem{worm} For a review of more recent approaches
to the same idea, see G. Gilbert, \npb{328}{1989}{159}
and references therein.

\bibitem{krauss} L.M. Krauss and F. Wilczek, 
\prd{62}{1989}{1221}.

\bibitem{anom} L.E. Ib\'a\~nez and G.G. Ross, 
\plb{260}{1991}{291}.

\bibitem{ross} L.E. Ib\'a\~nez and G.G. Ross,
\npb{368}{1992}{3}.

\bibitem{dine} T. Banks and M. Dine, 
\prd{45}{1992}{1424}.

\bibitem{ibanez} L.E. Ib\'a\~nez, \npb{398}{1993}{301}.

\bibitem{original}  K.R. Dienes, E. Dudas and T. Gherghetta,
\npb{557}{1999}{25}; N. Arkani-Hamed, S. Dimopoulos, G. Dvali
and J. March-Russell, \prd{65}{2002}{024032}.

\bibitem{sacha} K. Benakli and S. Davidson, \prd{60}{1999}{025004}.

\bibitem{gs} M. Green and J. Schwarz, \plb{149}{1984}{117}.

\bibitem{IRU98} L.E. Ib\'a\~nez, R. Rabad\'an, A.M. Uranga, \npb{542}{1999}{112}.

\bibitem{AFIRU00} G. Aldazabal, S. Franco, L.E. Ib\'a\~nez, R. Rabad\'an, A.M. Uranga,
\jhep{02}{2001}{047}.

\bibitem{IMR01} L.E. Ib\'a\~nez, F. Marchesano, R. Rabad\'an, \jhep{11}{2001}{002}.

\bibitem{BIQ98} C.P. Burgess, L.E. Ib\'a\~nez, F. Quevedo, \plb{477}{1999}{257}.

\bibitem{IQ99} L.E. Ib\'a\~nez, F. Quevedo, \jhep{10}{1999}{001}.

\bibitem{coleman} S. Coleman, J. Preskill, F. Wilczek, \npb{378}{1992}{175}.

\bibitem{GIIQ02} For a helpful discussion see:
D.M. Ghilencea, L.E. Ib\'a\~nez, N. Irges, F. Quevedo, \jhep{08}{2002}{016}.

\bibitem{newdine} M. Dine, M.L. Graesser, hep-th/0409209.

\end{thebibliography}
\end{document}